# Imaging the Renner-Teller effect using laser-induced electron diffraction

Kasra Amini[1,2,*], Michele Sclafani[1,*], Tobias Steinle[1,*], Anh-Thu Le[3,4], Aurelien Sanchez[1], Carolin Müller[5], Johannes Steinmetzer[5], Lun Yue[5], José Ramón Martínez Saavedra[1], Michaël Hemmer[1], Maciej Lewenstein[1,6], Robert Moshammer[7], Thomas Pfeifer[7], Michael G. Pullen[1], Joachim Ullrich[7,8], Benjamin Wolter[1], Robert Moszynski[2], F. Javier García de Abajo[1,6], C.D. Lin[3], Stefanie Gräfe[5,9], Jens Biegert[1,6,†]

[1]ICFO - Institut de Ciencies Fotoniques, The Barcelona Institute of Science and Technology, 08860 Barcelona, Spain.
[2]Department of Chemistry, University of Warsaw, Warsaw, 02-093, Poland.
[3]Department of Physics, J. R. Macdonald Laboratory, Kansas State University, Manhattan, KS, 66506-2604, USA.
[4]Department of Physics, Missouri University of Science and Technology, Rolla, MO 65409, USA.
[5]Institute of Physical Chemistry, Friedrich-Schiller University, Jena, 07743, Germany.
[6]ICREA, Pg. Lluís Companys 23, 08010 Barcelona, Spain.
[7]Max-Planck-Institute für Kernphysik, Heidelberg, 69117, Germany.
[8]Physikalisch-Technische Bundesanstalt (PTB), Braunschweig, D-38116, Germany.
[9]Abbe Center of Photonics, Friedrich-Schiller University, Jena, 07745, Germany.
*These authors contributed equally to this work.
†To whom correspondence should be addressed to. Email: jens.biegert@icfo.eu.

**Structural information on electronically excited neutral molecules can be indirectly retrieved, largely through pump-probe and rotational spectroscopy measurements with the aid of calculations. Here, we demonstrate the direct structural retrieval of neutral carbonyl disulfide ($CS_2$) in the $\widetilde{B}^1B_2$ excited electronic state using laser-induced electron diffraction (LIED). We unambiguously identify the ultrafast symmetric stretching and bending of the field-dressed neutral $CS_2$ molecule with combined picometre and attosecond resolution using intra-pulse pump-probe excitation and measurement. We invoke the Renner-Teller effect to populate the $\widetilde{B}^1B_2$ excited state in neutral $CS_2$, leading to bending and stretching of the molecule. Our results demonstrate the sensitivity of LIED in retrieving the geometric structure of $CS_2$, which is known to appear as a two-centre scatterer.**

### Significance

**Laser-induced electron diffraction is a molecular-scale electron microscope that captures clean snapshots of a molecule's geometry with sub-atomic picometre and attosecond spatio-temporal resolution. We induce and unambiguously identify the stretching and bending of a linear triatomic molecule following the excitation of the molecule to an excited electronic state with a bent and stretched geometry. We show that we can directly retrieve the structure of electronically excited molecules that is otherwise possible through indirect retrieval methods such as pump-probe and rotational spectroscopy measurements.**

Many important phenomena in biology, chemistry and physics can only be described beyond the Born-Oppenheimer (BO) approximation, giving rise to non-adiabatic dynamics and the coupling of nuclear (vibrational and rotational) and electronic motion in molecules (1-7). One prominent example where the BO approximation breaks down is the Renner-Teller effect (8, 9): in any highly symmetric linear molecule with symmetry-induced degeneracy of electronic states, non-adiabatic coupling of (vibrational) nuclear and electronic degrees of freedom can lead to the distortion of the nuclear framework on a timescale comparable with electronic motion. The system's symmetry is then reduced by the bending of the molecule to split the degenerate electronic state into two distinct potential energy surfaces (PESs), leading to a more stable, bent conformer.

Here, we demonstrate the direct imaging of Renner-Teller non-adiabatic vibronic dynamics in neutral $CS_2$ with combined picometre and attosecond resolution through intra-pulse pump-probe excitation and measurement with laser-induced electron diffraction (LIED) (10-16). Our results shed light on the vibronic excitation of a neutral linear molecule in the rising edge of our laser field that causes bending and stretching of the molecule. High momentum transfers experienced by the electron wave packet (EWP; $U_p$ = 85 eV) with large scattering angles enable the electron to penetrate deep into the atomic cores, allowing us to resolve a strongly symmetrically stretched and bent $CS_2$ molecule most likely in the $\widetilde{B}^1B_2$ excited electronic state.

Specifically, we pump and probe $CS_2$ molecules in a one-pulse LIED measurement to capture a single high-resolution snapshot of the molecular structure at around the peak of the strong laser field. By analyzing the angular dependence of the experimentally detected molecular interference signal, we directly retrieve a symmetrically stretched and bent $CS_2^+$ structure. We subsequently present results from state-of-the-art quantum dynamical calculations to investigate the mechanism behind the linear-to-bent transition that occurs in field-dressed $CS_2$.

**Molecular Structure Extraction**

Fig. 1 displays the results for three different electron returning energies, $E_R$ = 160, 170 and 180 eV. From the measured momentum distribution, shown in Fig. 1*A*, the molecular differential cross-section (DCS) weighted by the molecular ionization rate and the alignment distribution is extracted using the quantitative rescattering (QRS) theory (see SI Appendix). Molecular structural information is then obtained from the field-free molecular DCS *via* the molecular contrast factor (MCF). Fig. 1*B* shows the experimental MCF (black dots) and the theoretical MCFs corresponding to the equilibrium geometric structure of the $\widetilde{X}^1\Sigma_g^+$ electronic ground state (orange trace) (9), the quasilinear geometry (green trace) (17, 18) and the geometric structure that theoretically agrees best with the experimentally measured structure (red trace). Overall, there is a good fit between the experimental MCF and the theoretical MCF that best fits the experimental data. An additional peak is observed in the experimental data between 7.5 and 8.0 Å$^{-1}$ in Fig. 1*B* that is not captured by our best-fit single-structure theoretical MCF, and is most likely due to a small contribution from another structure. Nevertheless, the single-structure fitting algorithm used in this work already agrees well with the experimental MCFs for a rather broad range of momentum transfer from around 5.5 to 9.5 Å$^{-1}$, thus we believe that the extracted bent structure is the dominant one. Retrieving this information at different returning electron kinetic energies yields consistent results with bent and symmetrically stretched neutral $CS_2$, as shown in Fig. 1*C*.

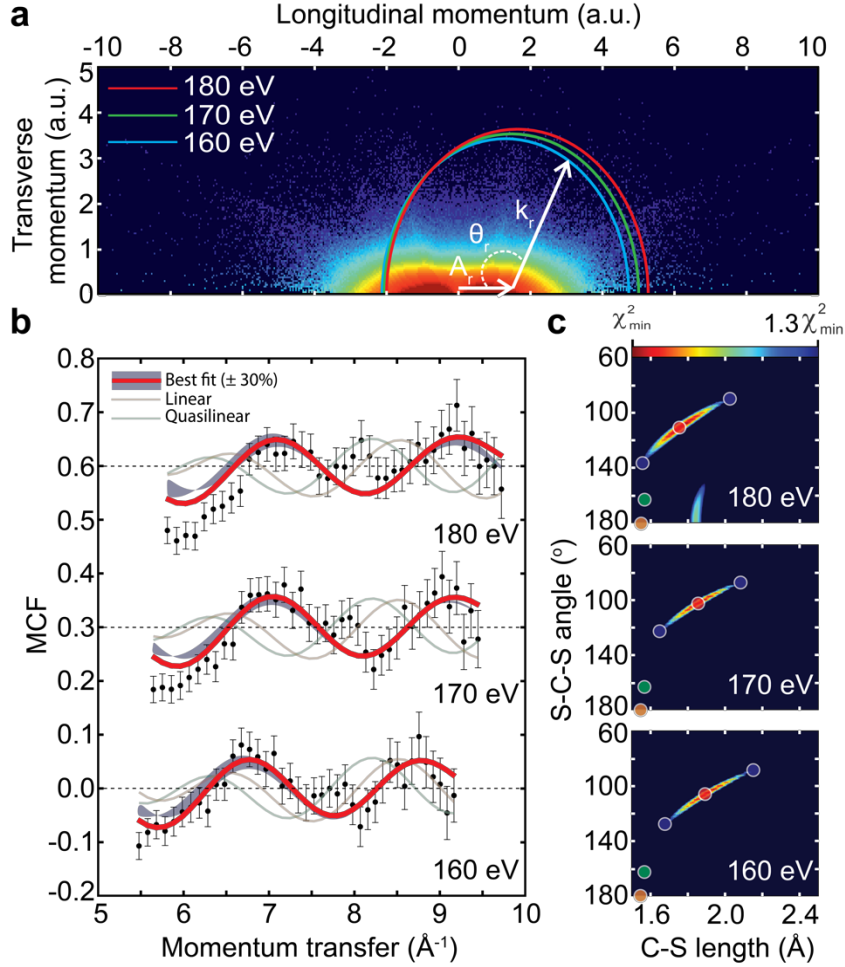

**Fig. 1 | LIED imaging of laser-induced skeletal deformations in CS$_2$.** (*A*) Double differential cross-sections are extracted by integrating the experimental momentum distribution map along the rescattering angle, θ$_r$, of the circle defined by the parametric relations $p_{long}$ = -$A_r$ ± ($k_r$ * cosθ$_r$) and $p_{trans}$ = $k_r$ * sinθ$_r$, where $A_r$ is the value of the field vector at the time of rescattering. (*B*) Comparison of the experimental (black dots) molecular contrast factor (MCF) to the theoretical MCFs associated with the equilibrium geometric structure of the $\widetilde{X}^1\Sigma_g^+$ electronic ground state (orange trace) (9), the quasilinear geometry (green trace) (17, 18) and the geometric structure that theoretically agrees best with the experimentally measured structure (red trace). The blue shaded region illustrates the sensitivity of the theoretical MCFs when varying $R_{CS}$ and $\Phi_{SCS}$ by around ±0.25 Å and ±20°, respectively, corresponding to a 30% increase from the χ² minimum (see SI Appendix). The data shown correspond to rescattered electrons with kinetic energies of 160, 170 and 180 eV. (*C*) CS$_2$ structural parameters are retrieved by locating the minimum of the χ² map (see SI Appendix, Eq. S1). Here, the most probable CS$_2$ geometry (red dot in each plot) is shown along with a 30% variation of the χ² minimum (blue dots). The orange circle indicates the equilibrium geometry of neutral CS$_2$ in its $\widetilde{X}^1\Sigma_g^+$ ground electronic state (1.55 Å, 180°) (9), whereas the green circle corresponds to CS$_2$ in a quasilinear configuration (1.54 Å, 163°) (17,18).

## Bent and Stretched Molecular Structure

The geometric parameters are retrieved from our LIED measurements as a function of the electron returning energy, as shown in Fig. 2. We measure a C-S bond length $R_{CS}$ = 1.86 ± 0.23 Å and an S-C-S angle $\Phi_{SCS}$ = 104.0 ± 20.2°, which correspond to a strongly symmetrically stretched and bent molecule. Since field-free neutral CS$_2$ in the ground electronic state, $\widetilde{X}^1\Sigma_g^+$, is linear in geometry ($R_{eq}$ = 1.55 Å and $\Phi_{SCS}$ = 180°) (18), a linear-to-bent transition occurs that leads to the experimentally measured bent LIED structure.

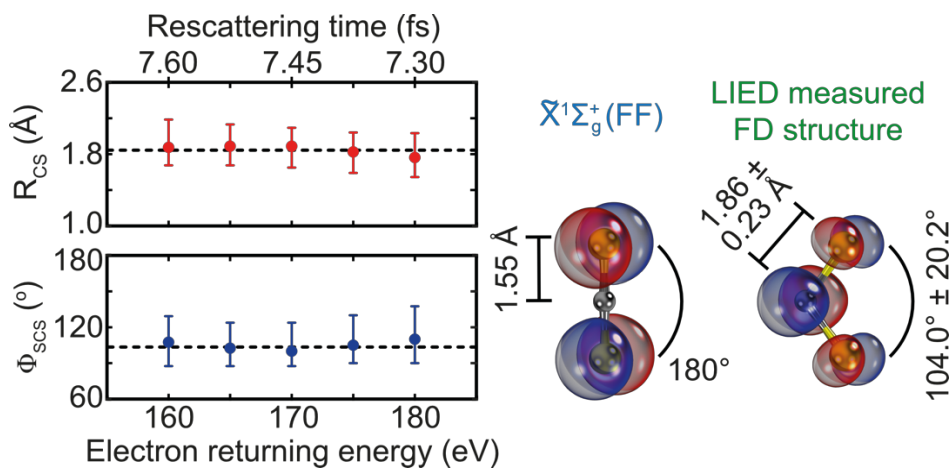

**Fig. 2 | Stretching and bending of field-dressed CS$_2$.** Geometrical parameters of CS$_2$ are retrieved as a function of the electron returning energy. By fitting a constant line, we estimate a C-S bond length $R_{CS}$ = 1.86 ± 0.23 Å and a S-C-S angle $\Phi_{SCS}$ = 104.0 ± 20.2°, which correspond to a strongly symmetrically stretched and bent neutral CS$_2$. The return time of the re-scattered electrons is also displayed at the top of the figure. Models with molecular orbitals are shown on the right-hand side of the figure for field-free (FF) neutral CS$_2$ in the ground electronic state, $\widetilde{X}^1\Sigma_g^+$, and the LIED measured field-dressed (FD) structure. The corresponding $R_{CS}$ and $\Phi_{SCS}$ values for these two structures are indicated.

## Quantum Chemistry Dynamical Calculations

We performed advanced, state-of-the-art quantum dynamical calculations of coupled electron-nuclear motions on the field-dressed PESs in the presence of an intense laser field to investigate the mechanism behind such a linear-to-bent transition (see SI Appendix). Our calculations reveal a Renner-Teller excitation mechanism that leads to the stretching and bending of neutral CS$_2$, with a schematic of the excitation shown in Fig. 3*A*.

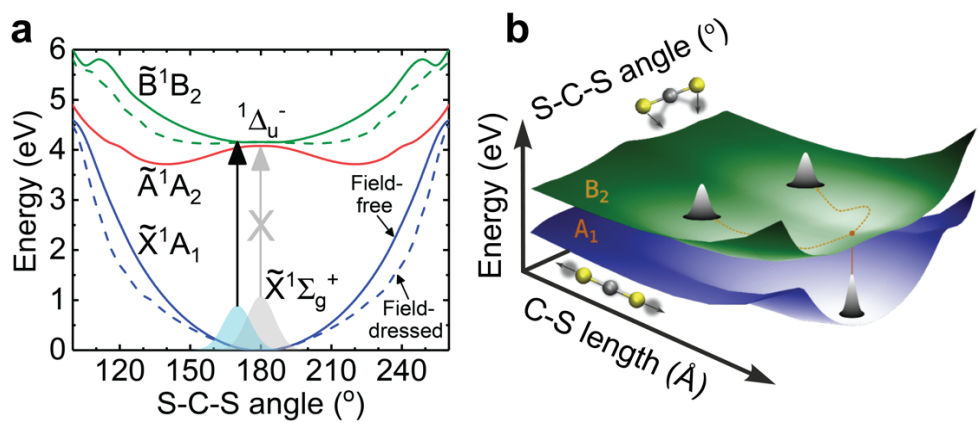

**Fig. 3 | Renner-Teller excitation mechanism in neutral CS$_2$.** (*A*) Potential energy curves (PECs) for the field-free (solid curves) neutral CS$_2$ in the ground electronic state, along with the $\widetilde{X}^1A_1$ (blue), the $\widetilde{A}^1A_2$ (red) and the $\widetilde{B}^1B_2$ (green) excited electronic states are shown as a function of the S-C-S angle at fixed $R_{CS}$ = 1.86 Å. The corresponding field-dressed (dashed curves) PECs are also shown. In the linear geometry (D$_{\infty h}$), a transition from the $\widetilde{X}^1\Sigma_g^+$ ground electronic state to the $^1\Delta_u$ excited electronic state is dipole forbidden (grey vertical arrow) due to symmetry considerations. However, our calculations show that the molecule begins to bend by 10° (C$_{2v}$) in the presence of a strong field. At the same time, at bent geometries, the two-fold degeneracy of $^1\Delta_u$ is lifted and splits into two distinct bent excited electronic states: $\widetilde{A}^1A_2$ and $\widetilde{B}^1B_2$. At these bent geometries, a transition from the $\widetilde{X}^1A_1$ ground state to the $\widetilde{B}^1B_2$ excited state becomes dipole allowed (black vertical arrow). (*B*) Potential energy surfaces (PESs) of field-dressed (FD) CS$_2$ in the $\widetilde{X}^1A_1$ ground electronic state and $\widetilde{B}^1B_2$ excited state. Once the $\widetilde{B}^1B_2$ state is populated, the nuclear wave packet evolves towards the equilibrium position of the $\widetilde{B}^1B_2$ state.

Optical excitation to the lowest-lying singlet excited electronic states, such as the doubly-degenerate $^1\Delta_u$ state, from the $\tilde{X}^1\Sigma_g^+$ ground state in field-free neutral $CS_2$ is strictly dipole-forbidden in the linear geometry ($D_{\infty h}$) due to symmetry considerations (grey arrow in Fig. 3A). However, in the presence of a strong field, our wave packet calculations in Fig. 4A show that the field-dressed (FD) molecule initially bends by approximately 10° within 90 fs (blue rectangle in Fig. 4A) to split the degeneracy of $^1\Delta_u$ into two bent states ($\tilde{A}^1A_2$ and $\tilde{B}^1B_2$) in neutral $CS_2$.

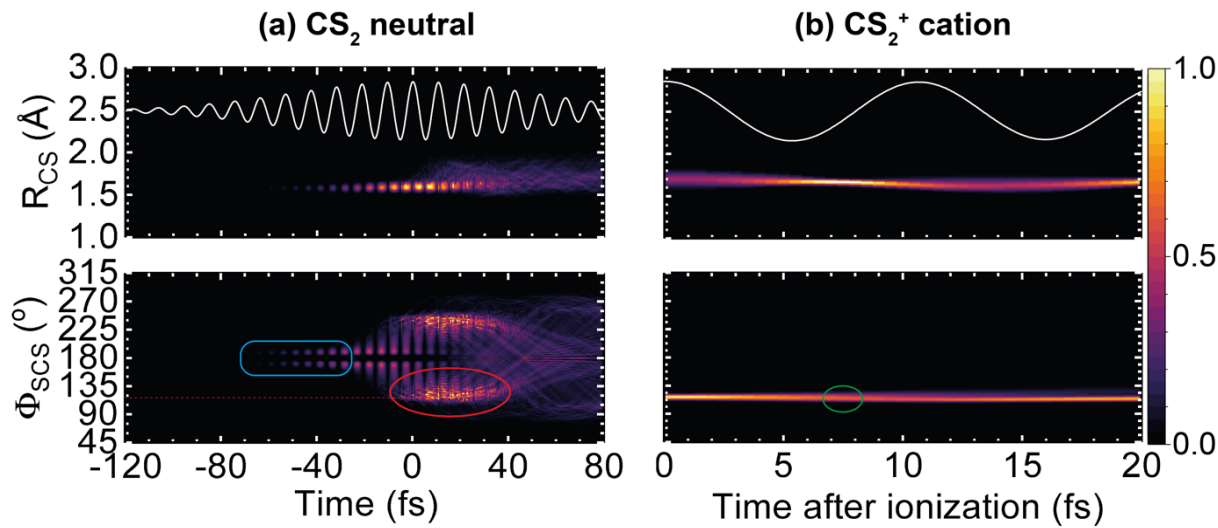

**Fig. 4 | Quantum dynamical wave packet calculations.** The stretching (*Top*) of C-S internuclear distance, $R_{CS}$, and bending (*Bottom*) of the S-C-S bond angle, $\phi_{SCS}$, for (A) neutral $CS_2$ in the $\tilde{B}^1B_2$ state and (B) $CS_2^+$ cation. The starting conditions used are: (A) neutral $CS_2$ in the $\tilde{X}^1\Sigma_g^+$ ground electronic state (1.55 Å, 180°); and (B) neutral $CS_2$ in the $\tilde{B}^1B_2$ excited electronic state (1.7 Å, 117°). The blue rectangle indicates the initial bending of neutral $CS_2$. The red (green) circle indicates the relevant structure at around the time of ionization (re-scattering), $t_i$ ($t_r$). Here, molecules are 90° to the laser polarization. In panel (*A*), $t$ = 0 fs corresponds to the peak of the 85 fs (FWHM) 3.1 μm pulse envelope, whilst in panel (*B*) the time axis corresponds to the time after ionization. The corresponding laser field is shown as white traces at the top of both panels.

This enables the nuclear wave packet to reach non-equilibrium positions in the initially bent molecule, such that only a transition from the $\tilde{X}^1A_1$ ground state to the $\tilde{B}^1B_2$ excited state becomes dipole-allowed (black arrow in Fig. 3A) in the bent geometry ($C_{2v}$). Our quantum dynamical calculations confirm that symmetric stretching and bending in the laser field occurs, leading to an estimated population of about 3% in the $\tilde{B}^1B_2$ state in neutral $CS_2$. Our calculations for neutral $CS_2$ in Fig. 4A show that the molecule in the excited state bends up to about 120° at $t$ = 0 fs (*i.e.* near the maximum of the pulse envelope; see red circle in Fig. 4A). The wave packet in the $\tilde{B}^1B_2$ state then proceeds to find its lowest-energy equilibrium position ($R_{eq}$ = 1.64 Å and $\phi_{SCS}$ = 130°) (16-19), as shown in Fig. 3B. Other excited electronic states are not populated due to small dipole couplings, even in the deformed geometry. Since the energy gap of $\tilde{B}^1B_2$ relative to the ground state is approximately 4.5 eV according to our calculations, the strong tunneling ionization from $\tilde{B}^1B_2$ completely dominates, which permits the identification of the $\tilde{B}^1B_2$ state. Moreover, our dynamical calculations also show that the geometry of the cation (1.74 Å, 102°) does not change significantly relative to the deformed excited neutral (1.70 Å, 117°) within half a laser cycle after tunnel ionization from the $\tilde{B}^1B_2$ state (*i.e.* during 7 – 8 fs excursion time of the re-scattering electron; see green circle in Fig. 4B).

The exact geometry of neutral $CS_2$ in the $\tilde{B}^1B_2$ excited electronic state is still discussed (19, 20); spectroscopic measurements by Jungen *et al.* (17) reported a quasilinear structure (1.544 ± 0.006 Å, 163°), while a much more recent analysis of the rotational progressions in the $\widetilde{B}^1B_2 \leftarrow \widetilde{X}^1\Sigma_g^+$ spectrum led to a largely corrected, significantly bent geometry (1.64 Å, 131.9°) (21). These measurements in fact indirectly retrieve structural information. Our directly measured structure (1.86 ± 0.23 Å, 104.0 ± 20.2°) is in general agreement with previous theoretical investigations (~1.64 Å,~130°) (18-20) into neutral $CS_2$ in the $\widetilde{B}^1B_2$ excited state. The MCF that corresponds to the quasilinear geometry previously measured (1.544 ± 0.006 Å, 163°) (17) does not agree with our measured data. In contrast, our results clearly support a symmetrically stretched and strongly bent molecular structure. Analogous observations of $CS_2$ skeletal deformation have been recently reported by Yang *et al.* (22), who imaged an increase in $R_{CS}$ by 0.16 Å and 0.20 Å with respect to the equilibrium bond length when a 60 fs, 800 nm laser pulse is increased in intensity from $1.3 \times 10^{13}$ Wcm$^{-2}$ to $2.4 \times 10^{13}$ Wcm$^{-2}$, respectively. An assumed linear extrapolation of their results would produce a 0.43 Å bond length increase for the intensity we use ($9 \times 10^{13}$ Wcm$^{-2}$), which is fully consistent with the value reported here of 0.31 ± 0.23 Å. This corresponds to a strongly symmetrically stretched C-S bonds in vibronically excited neutral $CS_2$. Although clear indications of symmetric bond elongation were observed by Yang *et al.* (22), no firm conclusion was drawn about the bending vibration because of the limited spatial resolution (1.2 Å) of their UED probe, due to the small momentum transfer of their scattered electrons (<3.5 Å$^{-1}$). It should also be noted that Yang *et al.* used a field-free probe of molecular structure through ultrafast electron diffraction (UED) with a ~400 fs pulse duration (22). Moreover, the lack of an electron-ion coincidence-based detection scheme added further ambiguity to the physical mechanism behind the IR-induced excitation, with two possible mechanisms suggested by the authors: excitation of an electronic state through a multiphoton process, and formation of ions with longer bond lengths.

We use LIED to directly retrieve the geometric transformation of neutral $CS_2$ due to the Renner-Teller effect. Our measurements unambiguously identify a bent and symmetrically stretched $CS_2$ molecule ($R_{CS}$ = 1.86 ± 0.23 Å, $\Phi_{SCS}$ = 104.0 ± 20.2°) that is most likely populating the $\widetilde{B}^1B_2$ excited electronic state. This finding is also supported by our state-of-the-art quantum dynamical *ab initio* molecular dynamics calculations, which describe the linear-to-bent $\widetilde{B}^1B_2 \leftarrow \widetilde{X}^1\Sigma_g^+$ transition in neutral $CS_2$. Moreover, previous theory and indirect measurements of neutral $CS_2$ in the $\widetilde{B}^1B_2$ excited state also broadly support our LIED measurement and calculations (18-21).

We find that the nuclear distortion in fact first proceeds through the stretching of the C-S bonds before the molecule departs from the linear geometry and begins to bend on the rising edge of the LIED pulse (at time $t_p$ in Fig. 5). Consequently, a bent neutral $CS_2$ molecule most likely in the $\widetilde{B}^1B_2$ excited electronic state is preferentially subsequently ionized at the peak of the pulse (at time $t_i$ in Fig. 5) to initiate the LIED process. LIED is the elastic rescattering of the highly-energetic returning EWP onto the molecular ion (at time $t_r$ in Fig. 5), with structural information embedded within the rescattered EWP's momentum distribution at the time of recollision (see Methods) (12, 14, 23). Here, the returning EWP scatters against the $CS_2^+$ molecular ion (at time $t_r$), which has a similar strongly stretched and bent geometry as the neutral $CS_2$ in an excited electronic state at the point of ionization (at time $t_i$ in Fig. 5). However, during the excursion time of the returning electron of about 7 - 8 fs, vibrational dynamics on the cationic potential energy curves in the presence of the laser field occurs. During that time, as our calculations show (see green circle in Fig. 4*B*), the excited cation bends slightly further leading to a structure that is in good agreement with the experimentally observed bent and stretched structure.

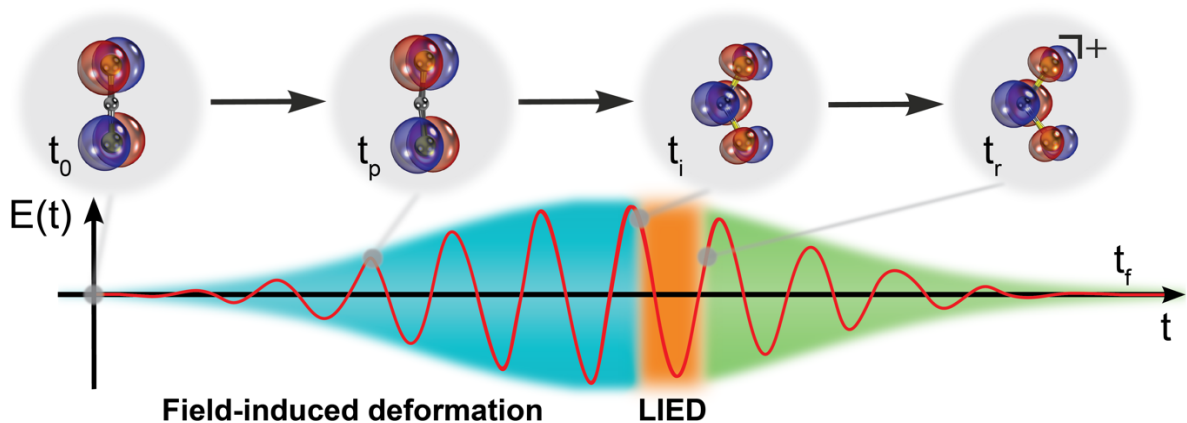

**Fig. 5 | Illustration of field-induced deformation and LIED measurement.** In our LIED measurement, the neutral $CS_2$ molecule is first symmetrically stretched and initially bent by 10° (at time $t_p$) before leading to the significantly bent $CS_2$ structure at the time of ionization, $t_i$. A high-resolution snapshot is recorded by the high energy electrons at the point of re-scattering, $t_r$.

Ultimately, our results illustrate the utility of intra-pulse LIED to retrieve structural transformation with combined picometre and attosecond resolution, allowing us to directly visualize non-adiabatic dynamics in molecular systems.

**Methods**

**A. Mid-IR OPCPA source.** A home-built optical parametric chirped pulse amplifier (OPCPA) set-up generates 85 fs, 3.1 μm pulses at a 160 kHz repetition rate with up to 21 W output power (24, 25). The OPCPA system is seeded by a passively carrier-envelope-phase (CEP) stable frequency comb generated by the difference frequency of a dual-colour fibre laser system (26). The mid-infrared wavelength of 3.1 μm ensures that the target is strong-field ionized in the tunnelling regime. The laser pulse is focused to a spot size of 6 – 7 μm resulting in a peak intensity of $9 \times 10^{13}$ Wcm$^{-2}$.

**B. ReMi detection system.** The experimental setup is based on a Reaction Microscope (ReMi) which has been previously described in detail in Refs. (27-29). Briefly, a doubly-skimmed supersonic jet of carbon disulfide provides the cold molecular target with a rotational temperature of <100 K. Homogeneous electric and magnetic extraction fields are employed to guide the ionic fragments and the corresponding electrons to separate detectors in the ReMi. Each detector consists of delay line detectors (Roentdek) which record the full three-dimensional momenta of charged particles from a single molecular fragmentation event in full electron-ion coincidence. In all experiments, the laser polarization is aligned perpendicular to the spectrometer axis, parallel to the jet.

**C. Molecular structure extraction.** Structural information of the molecular sample are retrieved from the electron momentum distribution within the frame of the quantitative rescattering theory (QRS) and the independent atomic-rescattering model (IAM) (30-32). We have extracted the molecular DCS from the experimental photoelectron momentum distribution as previously described in Ref. (14). See SI Appendix for further details.

**ACKNOWLEDGEMENTS.** We thank A. Stolow and J. Küpper for helpful and inspiring discussions. We acknowledge financial support from the Spanish Ministry of Economy and Competitiveness (MINECO), through the "Severo Ochoa" Programme for Centres of Excellence in R&D (SEV-2015-0522) Fundació Cellex Barcelona and the CERCA Programme / Generalitat de Catalunya. K.A., M.S., T.S., A.S., M.H., M.G.P., B.W. and J.B. acknowledge the European Research Council for ERC Advanced Grant TRANSFORMER (788218), MINECO for Plan Nacional FIS2017-89536-P; AGAUR for 2017 SGR1639, Laserlab-Europe (EU-H2020 654148). K.A., J.B., M.L. and R.M. acknowledge the Polish National Science Center within the project Symfonia, 2016/20/W/ST4/00314. A.S. and J.B. acknowledge Marie Sklodowska-Curie grant agreement No. 641272. F.J.G.A. acknowledges help from MINECO (MAT2017-88492-R) and ERC (Advanced Grant 789104-eNANO). C. M. and S. G. highly acknowledge the ERC Consolidator Grant QUEMCHEM (772676). L. Y. and S. G. acknowledge funding from the German Research Foundation, Grant number GR 4482/2. A.T.L. and C.D.L. are supported by the U.S. Department of Energy (DOE) under Grant No. DE-FG02-86ER13491. M.L. acknowledges support from the Ministerio de Economía y Competitividad through Plan Nacional (Grant No. FIS2016-79508-P FISICATEAMO), de Catalunya (Grant No. SGR 1341), CERCA Programme, the ERC (Advanced Grant OSYRIS), and funding through EU FETPRO QUIC.

Author contributions: J.B. designed research; K.A., M.S., T.S., A.T.L., A.S., C.M., J.S., L.Y., J.R.M.S., M.H., M.L., R. Moshammer, M.G.P., J.U., B.W., and J.B. performed research; A.T.L., C.M., J.S., L.Y., J.R.M.S., R. Moszynski, F.J.G.d.A., C.D.L., and S.G. contributed new reagents/analytic tools; K.A., M.S., T.S., A.S., and M.G.P. analyzed data; and K.A., M.S., T.S., A.T.L., M.L., R. Moshammer, T.P., J.U., R. Moszynski, F.J.G.d.A., C.D.L., S.G., and J.B. wrote the paper.

The authors declare no competing financial interests.